\documentclass[epj]{svjour} 
\journalname{}

\usepackage{amsmath}
\usepackage{amssymb}
\usepackage[margin=2cm]{geometry}
\usepackage{subfigure}
\usepackage{graphicx}
\usepackage{cite}
\usepackage{here}

\def\q<#1>{\langle #1 \rangle_q}%
\def\qone<#1>{\langle #1 \rangle_{q=1}}%
\def\<#1>{\langle #1 \rangle}%

\newcommand\Tr{\mathrm{Tr}}
\newcommand\betaph{\beta_{\mathrm{ph}}}
\newcommand\Tph{T_{\mathrm{ph}}}
\newcommand\Vpot{V_{\mathrm{pot}}}
\newcommand\VpotHF{V_{\mathrm{pot}}^{\mathrm{HF}}} 
\newcommand\VpotHFR{V_{\mathrm{pot}}^{\mathrm{HF,R}}} 
\newcommand\phic{\phi_{\mathrm{c}}}
\newcommand\phiic{\phi_{i\mathrm{c}}}
\newcommand\phijc{\phi_{j\mathrm{c}}}
\newcommand\UHF{U^{\mathrm{HF}}}
\newcommand\HHF{H^{\mathrm{HF}}}
\newcommand\phicVAC[1]{\overline{\phi}_{#1\mathrm{c}}}
\newcommand\massVAC[1]{\overline{m}_{#1}}
\newcommand\Ueff{U^{\mathrm{eff}}}

\newcommand\qexpectfree[1]{ \langle #1 \rangle_q^{(\beta), f}} 
\newcommand\qoneexpectfreephys[1]{ \langle #1 \rangle_{q=1}^{(\beta_{\mathrm{ph}}), f}}
\newcommand\expectremoval[1]{\langle\hspace{-0.2em}\langle #1 \rangle\hspace{-0.2em}\rangle} 

\begin{document}
\title{Chiral phase transition in the linear sigma model within Hartree factorization in the Tsallis nonextensive statistics}
\titlerunning{Chiral phase transition in the linear sigma model  in the Tsallis noextensive statistics}
\author{Masamichi Ishihara \thanks{\email{m\_isihar@koriyama-kgc.ac.jp}}}
\authorrunning{M.~Ishihara}
\institute{Department of Human Life Studies, Koriyama Women's University, Koriyama, Fukushima, 963-8503, Japan}

\abstract{
We studied chiral phase transition in the linear sigma model within the Tsallis nonextensive statistics
in the case of small deviation from the Boltzmann-Gibbs (BG) statistics. 
The statistics has two parameters: the temperature $T$ and the entropic parameter $q$. 
The normalized $q$-expectation value and the physical temperature $\Tph$ were employed in this study. 
The normalized $q$-expectation value was expanded as a series of the value $(1-q)$, 
where the absolute value $|1-q|$ is the measure of the deviation from the BG statistics.
We applied the Hartree factorization and the free particle approximation,  
and obtained the equations for the condensate, the sigma mass, and the pion mass. 
The physical temperature dependences of these quantities were obtained numerically. 
We found following facts.
The condensate at $q$ is smaller than that at $q'$ for $q>q'$. 
The sigma mass at $q$ is lighter than that at $q'$ for $q>q'$ at low physical temperature, 
and the sigma mass at $q$ is heavier than that at $q'$ for $q>q'$ at high physical temperature. 
The pion mass  at $q$ is heavier than that at $q'$ for $q>q'$.
The difference between the pion masses at different values of $q$ is small for $\Tph \le 200$ MeV.
That is to say,
the condensate and the sigma mass are affected by the Tsallis nonextensive statistics of small $|1-q|$,  
and the pion mass is also affected by the statistics of small $|1-q|$ except for $\Tph \le 200$ MeV. 
} 
\maketitle

\section{Introduction}

Power-like distributions have been interested by many researchers.
A statistics called the Tsallis nonextensive statistics \cite{Book:Tsallis} was proposed to describe power-like distributions.
The Tsallis nonextensive statistics has two parameters: the temperature $T$ and the entropic parameter $q$, 
and the deviation from the Boltzmann-Gibbs statistics is measured with $|1-q|$. 
Some definitions of the expectation value were proposed in the statistics \cite{Tsallis1998}.   
One of them is called normalized $q$-expectation value. 
The physical temperature $\Tph$ 
\cite{Book:Tsallis, Kalyana2000, Abe-PLA2001, Aragao-PhysicaA2003, Eicke-prepri2003, Toral-PhysicaA2003, Suyari-PTPsupple2006}
is often used to describe phenomena in the statistics. 
Power-like nature has been studied within the framework of the Tsallis nonextensive statistics.

Power-like distributions have been used to describe momentum distributions in high energy collisions.
Many researchers have studied momentum distributions, 
and have used Tsallis-type distributions
\cite{Alberico2009, Urmossy2011-PLB701, Cleymans2012, Cleymans2013-PLB723, Marques2015, GS2015, Azmi2015, Zheng2016, Thakur-AHEP2016, Lao2017, Cleymans2017-WoC, Cleymans2017, Yin2017, Osada-Ishihara-2018, Bhattacharyya2017-JPhysG45, Si-AHEP2018, Shen2018-PhysicaA492, Osada2019-Prepri}. 
It has been reported that Tsallis-type distributions describe well momentum distributions.
The value of $q$ was estimated in these studies. 
The deviation from the Boltzmann-Gibbs statistics is small at high energies, 
and the values of $|1-q|$ are close to $0.1$.
The effects of the small deviation from the Boltzmann-Gibbs statistics on physical quantities, 
such as correlation and fluctuation \cite{Ishihara2017-1, Ishihara2017-2, Ishihara2018, Osada-Ishihara-2018, Osada2019-Prepri}, have been studied.

Chiral symmetry restoration is an interesting topic in high energy heavy ion collisions, 
and it is believed that the symmetry is restored at high energies. 
The expectation value of a physical quantity, such as the square of a field, is affected by momentum distributions.
The effective potential depends on momentum distributions, and the symmetry restoration is also affected by momentum distributions. 
Therefore, the power-like distribution like the Tsallis distribution may change the values related to the chiral symmetry: the condensate and the masses. 
The phase transition of chiral symmetry is an attractive topic in the Tsallis nonextensive statistics
\cite{Rozynek2009, Ishihara2015, Rozynek2016, Ishihara2016, Shen2017, Ishihara2019}.

In this paper, we studied the physical temperature dependences of the condensate, the sigma mass, and the pion mass
for various $q$ in the linear sigma model within the framework of the Tsallis nonextensive statistics of small $|1-q|$.
We adopted the normalized $q$-expectation value.
We applied the Hartree factorization and 
the free particle approximation that the Hamiltonian in the density operator is  replaced with the free Hamiltonian. 
We calculated the physical temperature dependences of the condensate, the sigma mass, and the pion mass for various $q$. 
We found that 
the chiral symmetry restoration at $q$ occurs at low physical temperature, compared with the restoration at $q'$, for $q>q'$. 
Moreover, the difference between the pion mass at $q$ and the pion mass at $q'$ is small at low physical temperature.

This paper is organized as follows.
In sect.~\ref{Sec:phasetransition}, 
we derive the gap equations in the linear sigma model in the Tsallis nonextensive statistics of small $|1-q|$.
The physical temperature is introduced, and the normalized $q$-expectation value of a quantity is expanded as a series of $(1-q)$. 
The equations for the condensate and the masses are derived in the Hartree factorization and the free particle approximation.
In sect.~\ref{Sec:NumericalCalculation},
the derived self-consistent equations are solved numerically. 
The physical temperature dependences of the condensate, the sigma mass, and the pion mass are calculated for various $q$. 
The last section is assigned for the discussion and conclusion. 

\section{Chiral phase transition in the linear sigma model in the Tsallis nonextensive statistics}
\label{Sec:phasetransition}
\subsection{Tsallis nonextensive statistics within $(1-q)$ expansion}
We begin with the brief introduction of the Tsallis nonextensive statistics. 
The density operator $\rho$ in the statistics is given by \cite{Book:Tsallis, Aragao-PhysicaA2003}
\begin{align}
\rho &:=  (Z_q)^{-1}  \ \rho_u , 
\\
& \rho_u = \left[ 1 - (1-q) \frac{\beta}{c_q} (H-\q<H>)\right]^{1/(1-q)} ,  
\nonumber \\
& Z_q := \Tr \rho_u ,  
\nonumber 
\end{align}
where 
$\beta$ is the inverse temperature, $q$ is the entropic parameter, $c_q$ is a constant, and 
$H$ is the Hamiltonian, $\q<H>$ is the normalized $q$-expectation value of the Hamiltonian.
The $q$-expectation value of a quantity $A$ in the Tsallis nonextensive statistics is defined by
\begin{align}
\q<A>^{(\beta)} := \frac{\Tr(\rho^q A)}{\Tr(\rho^q)} . 
\end{align}
The constant $c_q$ is related to the partition function $Z_q$:
\begin{align}
c_q = (Z_q)^{1-q} . 
\label{cqZq}
\end{align} 

The $q$-dependent values are expanded as series of $\varepsilon \equiv 1-q$ to proceed the calculation
\cite{Ishihara2019, Ishihara2018}. 
\begin{subequations}
\begin{align}
&\q<H>^{(\beta)} = E_{[0]} - \varepsilon E_{[1]} + O(\varepsilon^2) \label{expand:E} ,\\
& c_q = c_{[0]} - \varepsilon c_{[1]} + O(\varepsilon^2) \label{expand:c} ,\\
& Z_q = Z_{[0]} - \varepsilon Z_{[1]} + O(\varepsilon^2)  \label{expand:Z} ,
\end{align}
\end{subequations}
where we attach the subscript $[j]$ to the coefficients of $\varepsilon^j$. 
It is obvious from eq.~\eqref{cqZq} that the constant $c_{[0]}$ is $1$.
We note that the coefficient $c_q$ is a function of $\beta$. 
We attach the superscript $\beta$ to $c_q$ as $c_q^{(\beta)}$:
 $c_q^{(\beta)} = 1 - \varepsilon c_{[1]}^{(\beta)} + O(\varepsilon^2)$.

The inverse physical temperature $\betaph$ is defined by 
\begin{align}
\betaph := \beta/c_q^{(\beta)}  . 
\end{align}
The inverse temperature $\beta$ is represented as a series of $\varepsilon$:
\begin{align}
\beta = \betaph - \varepsilon c_{[1]}^{(\betaph)} \betaph + O(\varepsilon^2) . 
\end{align}

The $(1-q)$ expansion of the normalized $q$-expectation value of the quantity $A$ \cite{Ishihara2018} is given by 
\begin{align}
\q<A>^{(\beta)}  &=  \qone<A>^{(\betaph)} 
+ \varepsilon 
\Big\{ \betaph\left( 1 + \betaph E_{[0]}^{(\betaph)} \right) 
    \nonumber \\ & \qquad 
    \times 
\left[\qone<HA>^{(\betaph)} - \qone<H>^{(\betaph)} \qone<A>^{(\betaph)} \right] 
\nonumber \\ & \quad 
- \frac{1}{2} \left( \betaph \right)^2 \left[\qone<H^2A>^{(\betaph)} - \qone<H^2>^{(\betaph)} \qone<A>^{(\betaph)} \right] \Big\} 
\nonumber \\ & \quad  
+ O(\varepsilon^2) . 
\label{final-eq-O-betaphys}
\end{align}

The number of the particles of a field $\varphi_s$ with the momentum $\vec{k}$ and the particle mass $m_s$ 
is given by substituting $A= n_{s\vec{k}} \equiv  a_{s\vec{k}}^{\dag} a_{s\vec{k}}$ into the previous equation, 
where $a_{j\vec{k}}^{\dag}$ and $a_{j\vec{k}}$ are creation and annihilation operators of a field $\varphi_j$ respectively, 
with the commutation relation $[a_{i\vec{k}}, a_{j\vec{l}}^{\dag}] = \delta_{i,j} \delta_{\vec{k}, \vec{l}}$. 
We employ the free Hamiltonian to calculate the number of the particles with momentum $\vec{k}$.
The free Hamiltonian $H^f$ of $N$-fields is 
\begin{align}
H^f &= \sum_{i=0}^{N-1} H_i^f , 
\label{eqn:free-Hamiltonian} \\ 
& H_i^f := \int d\vec{x} \left[ \frac{1}{2} (\partial^0 \varphi_i)^2 +  \frac{1}{2} (\nabla \varphi_i)^2  + \frac{1}{2} m_i^2 (\varphi_i)^2 \right] . 
\nonumber 
\end{align}

The number of the particles with the momentum $\vec{k}$ for the field $\varphi_s$ in the case of the free Hamiltonian is given by \cite{Ishihara2019}:
\begin{subequations}
\begin{align}
& \qexpectfree{n_{s\vec{k}}}  
 \nonumber \\ & 
 = \qoneexpectfreephys{n_{s\vec{k}}} 
 \nonumber \\ & \quad 
+ \varepsilon \Big\{ (\betaph \omega_{s\vec{k}} )\left( 1 + ( \betaph \omega_{s\vec{k}} ) \qoneexpectfreephys{n_{s\vec{k}}} \right) 
 \nonumber \\ & \qquad \times 
    \left[\qoneexpectfreephys{(n_{s\vec{k}})^2 } - \left( \qoneexpectfreephys{n_{s\vec{k}}} \right)^2 \right] 
\nonumber \\ &  \quad 
- \frac{1}{2} \left( \betaph \omega_{s\vec{k}} \right)^2 
\Big[\qoneexpectfreephys{(n_{s\vec{k}})^3} 
   \nonumber \\ & \qquad 
- \qoneexpectfreephys{(n_{s\vec{k}})^2} \qoneexpectfreephys{n_{s\vec{k}}} \Big] \Big\} 
 \nonumber \\ & \quad 
+ O(\varepsilon^2) 
\label{eq:nsk-freeHamiltonian-betaphys}
\\
&= \frac{1}{\exp(\betaph \omega_{s\vec{k}}) -1} 
+ \varepsilon \Bigg\{ 
        \frac{ (\betaph \omega_{s\vec{k}})  \exp(\betaph \omega_{s\vec{k}}) }{\left[ \exp(\betaph \omega_{s\vec{k}}) -1 \right]^2} 
 \nonumber \\ & \quad 
       - \frac{(\betaph \omega_{s\vec{k}})^2 \exp(\betaph \omega_{s\vec{k}}) \left[ \exp(\betaph \omega_{s\vec{k}}) +1 \right] }
             {2 \left[ \exp(\betaph \omega_{s\vec{k}}) -1 \right]^3} 
        \Bigg\}
\nonumber \\ & \qquad 
+ O(\varepsilon^2) 
,
\end{align}
\label{eq:nsk-freeHamiltonian-betaphys:all}
\end{subequations}
where we attach the superscript $f$ to clarify that the free Hamiltonian is adopted in the density operator.

The $(1-q)$ expansion of the normalized $q$-expectation value $\qexpectfree{(\varphi_j)^2}$ to the $O(\varepsilon)$
is obtained  by using eq.~\eqref{eq:nsk-freeHamiltonian-betaphys:all}:

\begin{align}
&\qexpectfree{(\varphi_j)^2} 
\nonumber \\  
&= 
\frac{1}{V} \sum_{\vec{k}} \left( \frac{1}{2 \omega_{j\vec{k}}} \right)  
 \nonumber \\ & \quad 
+ 
\frac{1}{V} \sum_{\vec{k}} \left( \frac{1}{\omega_{j\vec{k}}} \right) 
\left\{ 1 + \varepsilon \left[ (\betaph \omega_{j\vec{k}}) - \frac{1}{2} (\betaph \omega_{j\vec{k}})^2  \right] \right\}  
      \nonumber \\ & \qquad \times
\frac{1}{\big[ \exp(\betaph \omega_{j\vec{k}}) -1 \big]} 
\nonumber \\ & \quad  + 
\frac{1}{V} \sum_{\vec{k}}  \left( \frac{1}{\omega_{j\vec{k}}} \right)
\varepsilon \left[ (\betaph \omega_{j\vec{k}}) - \frac{3}{2} (\betaph \omega_{j\vec{k}})^2  \right] 
     \nonumber \\ & \qquad \times
\frac{1}{\big[ \exp(\betaph \omega_{j\vec{k}}) -1 \big]^2} 
\nonumber \\ & \quad  + 
\frac{1}{V} \sum_{\vec{k}} \left( \frac{1}{\omega_{j\vec{k}}} \right)
(-\varepsilon) (\betaph \omega_{j\vec{k}})^2  
\frac{1}{\big[ \exp(\betaph \omega_{j\vec{k}}) -1 \big]^3} 
  \nonumber \\ & \quad  
+ O(\varepsilon^2) . 
\label{eqn:freeHamiltoninan:varphi2}
\end{align}
We define the following function $F(x; a, b)$ to simplify the expression of $\qexpectfree{(\varphi_j)^2}$:
\begin{align}
F(x; a, b) :=  \int_0^{\infty} dy\  y^2 \frac{(y^2+x^2)^{a/2}}{\big[ \exp(\sqrt{y^2+x^2}) -1 \big]^b}  .
\end{align}
We also define the variable $x_j$ by 
\begin{align}
x_j := \betaph m_j. 
\end{align}
The quantity $\qexpectfree{(\varphi_j)^2}$ is expressed with the function $F(x; a, b)$ as $V$ approaches infinity. 
\begin{align}
& 
\qexpectfree{(\varphi_j)^2} 
\nonumber \\ 
&= \frac{1}{4\pi^2 (\betaph)^2} F(x_j,-1,0) 
+ \frac{1}{2\pi^2 (\betaph)^2} \Bigg\{ F(x_j,-1,1) 
\nonumber \\ & \quad 
+  \varepsilon\Bigg[ F(x_j; 0,1) - \frac{1}{2} F(x_j; 1,1) + F(x_j; 0,2) 
\nonumber \\ & \qquad 
- \frac{3}{2} F(x_j; 1,2) - F(x_j; 1,3) \Bigg]
\Bigg\}
+O(\varepsilon^2) .
\label{varphi2:explicit}
\end{align}
The above result is used in the next subsection.


\subsection{Gap equations in Hartree factorization}
In this study, we use the linear sigma model to study the chiral phase transition.
The Hamiltonian $H$ of the linear sigma model is 
\begin{subequations}
\begin{align}
&H = \int d\vec{x} \left[ \frac{1}{2} (\partial^0 \phi)^2  + \frac{1}{2} (\nabla \phi)^2  \right] + \int d\vec{x}  V_{\mathrm{pot}} (\phi) , \\
& V_{\mathrm{pot}} (\phi)   = \frac{\lambda}{4} (\phi^2 - v^2)^2 - G \phi_0, 
\end{align}
\end{subequations}
where $\phi \equiv (\phi_0, \phi_1,  \cdots, \phi_{N-1})$, 
$\phi^2 \equiv \sum_{j=0}^{N-1} \phi_j^2$, and  $(\partial^i \phi)^2 \equiv \sum_{j=0}^{N-1} (\partial^i \phi_j)^2$.
The quantities, $\lambda$, $v$, and $G$, are the parameters of the linear sigma model.
We shift the fields as $\phi_j = \phi_{j\mathrm{c}} + \varphi_j$,
where the quantity $\phi_{j\mathrm{c}}$ is the condensate of the field $\phi_j$. 
The Hamiltonian is given by 
\begin{subequations}
\begin{align}
H &= \int d\vec{x} \sum_j \left[ \frac{1}{2} (\partial^0 \varphi_j)^2  + \frac{1}{2} (\nabla \varphi_j)^2  \right] 
\nonumber \\ & \quad 
           + \int d\vec{x}  \Vpot  (\phi_j = \phijc + \varphi_j)  \\
    & = \int d\vec{x}  \sum_j  \left[ \frac{1}{2} (\partial^0 \varphi_j)^2  + \frac{1}{2} (\nabla \varphi_j)^2  + \frac{1}{2} m_j^2 \varphi_j^2\right] 
\nonumber \\ & \quad 
            + \int d\vec{x}  \Vpot (\phi_j = \phijc + \varphi_j)  -  \int d\vec{x}  \sum_j \left( \frac{1}{2} m_j^2 \varphi_j^2 \right)
.
\end{align}
\end{subequations}
where it is assumed that $\phi_{j \mathrm{c}}$ is uniform and independent of  time.  
The mass $m_j$ for the field $\varphi_j$ is discussed later. 

The Hartree factorization \cite{Boyanovsky:PRD55, Boyanovsky:PRD56} is applied to the above Hamiltonian, 
and the potential term $V_{\mathrm{pot}}$ in the Hartree factorization is given by
\begin{align}
& 
\VpotHF (\phi_j = \phijc + \varphi_j) 
\nonumber \\ 
&= \frac{\lambda}{4} \left\{ (\phic^2 - v^2)^2 - (\<\varphi^2>)^2 - 2 \sum_{i=0}^{N-1} \left( \<\varphi_i^2> \right)^2 \right\}
    - G \phi_{0\mathrm{c}}  
\nonumber \\  &  \quad
   +\frac{\lambda}{4} \Bigg\{ \left[ 2 ( \phic^2 - v^2) + 4 \< \varphi^2> \right] ( \phic \cdot \varphi) 
 \nonumber \\  & \qquad   
   + 8 \sum_{i=0}^{N-1} \phiic \< \varphi_i^2 > \varphi_i \Bigg\}  - G \varphi_0
\nonumber \\  & \quad   
   + \frac{\lambda}{4} \Bigg\{ 
   \left[ 2 ( \phic^2 - v^2) + 2 \< \varphi^2> \right]  \varphi^2 + 4  ( \phic \cdot \varphi)^2 
 \nonumber \\  & \qquad   
   + 4 \sum_{i=0}^{N-1} \< \varphi_i^2 > \varphi_i^2 
   \Bigg\} 
. 
\end{align}
where the representation $(\phic \cdot \varphi)$ indicates $\sum_{j=0}^{N-1} \phijc \varphi_j$.
The notation $\<O>$ is the expectation value of a quantity $O$. 
The Hamiltonian in the Hartree factorization is given by
\begin{align}
\HHF &= \int d\vec{x}  \sum_j  \left[ \frac{1}{2} (\partial^0 \varphi_j)^2  + \frac{1}{2} (\nabla \varphi_j)^2  + \frac{1}{2} m_j^2 \varphi_j^2\right] 
\nonumber \\ &  \qquad 
 + \int d\vec{x} \ \VpotHF(\phi_j = \phijc + \varphi_j)  
\nonumber \\ &  \qquad 
- \int d\vec{x}\  \sum_{j} \left( \frac{1}{2} m_j^2 \varphi_j^2 \right) .
\end{align}

We apply the free particle approximation that the Hamiltonian is replaced with $H^f$, eq.~\eqref{eqn:free-Hamiltonian}, in the density operator.
The potential $\UHF(\{\phijc\},\{m_j\})$ with the normalized $q$-expectation value in this approximation is defined by 
\begin{align} 
\UHF(\{\phijc\},\{m_j\}) := V^{-1} \qexpectfree{\HHF}. 
\end{align}
The notations $\{\phijc\}$ and $\{m_j\}$ represent the set of $\phijc$ and the set of $m_j$, respectively.

The potential $\UHF$ is represented with the annihilation operator $a_{j\vec{k}}$ for the field $\varphi_j$ and the energy $\omega_{j\vec{k}}$:
\begin{align} 
& 
\UHF(\{\phijc\},\{m_j\}) 
\nonumber \\ & 
= V^{-1} \Bigg\{ 
\sum_j \sum_{\vec{k}} \omega_{j\vec{k}} \Big[ \qexpectfree{a_{j\vec{k}}^{\dag} a_{j\vec{k}} } + \frac{1}{2} \Big] 
 \nonumber \\ & \quad 
+ \int d\vec{x} \ \qexpectfree{\VpotHF} - \sum_{j} \frac{1}{2} m_j^2 \int d\vec{x} \qexpectfree{\varphi_j^2} 
\Bigg\} 
.
\label{eqn:IHF}
\end{align}

We use the mass $m_j$ on the energy minimum state (vacuum).   
For the field $\phi_j$, we denote the condensate on the vacuum as $\phicVAC{j}$
and denote the mass on the vacuum as $\massVAC{j}$.
We note that the mass $\massVAC{j}$ is constant at a fixed temperature $T$. 
That is, the mass $\massVAC{j}$ is not a function of $\phijc$.

The effective potential $\Ueff$ is defined by
\begin{align}
\Ueff(\{\phijc\}) := \UHF(\{\phijc\}, \{\massVAC{j}\}) .
\end{align}
The following conditions are adopted to determine the condensates and masses.
\begin{align}
\frac{\partial}{\partial \phi_{s\mathrm{c}}} \Ueff(\{\phijc\}) & = 
\left. \frac{\partial}{\partial \phi_{s\mathrm{c}}} \UHF(\{\phijc \}, \{m_{j}\})  \right|_{m_j = \massVAC{j}} 
= 0 
\nonumber \\ & \qquad 
\qquad (s=0,1, \cdots N-1)
.
\label{eqn:cond:condensate}
\end{align}
The  condensation $\phicVAC{j}$ is  the solution of eq.~\eqref{eqn:cond:condensate}.
In the present case, the potential is tilted to the $j=0$ direction,  we set $\phicVAC{j}=0$ for $j \neq 0$. 
Therefore, eq.~\eqref{eqn:cond:condensate} is reduced to 
\begin{align}
\left. 
\frac{\partial}{\partial \phi_{0\mathrm{c}}} \Ueff(\{\phijc \}) 
\right|_{\{\phi_{0\mathrm{c}} = \phicVAC{0}, \phi_{1\mathrm{c}}=0, \cdots, \phi_{(N-1)\mathrm{c}}=0\}} 
 = 0 
. 
\label{eqn:cond:phi0c}
\end{align}
That is, the condensate $\phicVAC{0}$ is given as the solution of eq.~\eqref{eqn:cond:phi0c}.
The condition for the mass $\massVAC{j}$ is 
\begin{align}
&\left. \frac{\partial^2}{\partial \phi_{s\mathrm{c}}^2} \Ueff(\{\phijc\})  \right|_{\{\phi_{0\mathrm{c}} = \phicVAC{0}, \phi_{1\mathrm{c}}=0, \cdots, \phi_{(N-1)\mathrm{c}}=0\}} 
\nonumber \\
&= \left. 
\frac{\partial^2}{\partial \phi_{s\mathrm{c}}^2} \UHF(\{\phijc\}, \{m_j\})  
\right|_{\substack{\{\phi_{0\mathrm{c}} = \phicVAC{0}, \phi_{1\mathrm{c}}=0, \cdots, \phi_{(N-1)\mathrm{c}}=0\}, \\ \{m_j=\massVAC{j}\} \hphantom{\phi_{1\mathrm{c}}=0, \cdots, \phi_{(N-1)\mathrm{c}}=0} }} 
\nonumber \\ &  
= \massVAC{s}^2
\qquad (s=0, 1,\cdots, N-1)
.
\label{eqn:cond:mass}
 \end{align}
Equations~\eqref{eqn:cond:phi0c} and \eqref{eqn:cond:mass} are rewritten as follows,
because the first and last terms in the brace of eq.~\eqref{eqn:IHF} do not include $\phijc$, 
while the terms include the masses $m_j$.
\begin{subequations}
\begin{align}
& \left. \frac{\partial}{\partial \phi_{0\mathrm{c}}} \Ueff(\{\phijc\})  \right|_{\{\phi_{0\mathrm{c}} = \phicVAC{0}, \phi_{1\mathrm{c}}=0, \cdots, \phi_{(N-1)\mathrm{c}}=0\}} 
\nonumber \\ &
= \left. \frac{1}{V} \frac{\partial}{\partial \phi_{0\mathrm{c}}} \int d\vec{x} \qexpectfree{\VpotHF}  
   \right|_{\substack{\{\phi_{0\mathrm{c}} = \phicVAC{0}, \phi_{1\mathrm{c}}=0, \cdots, \phi_{(N-1)\mathrm{c}}=0\}, \\ \{m_j=\massVAC{j}\} \hphantom{\phi_{1\mathrm{c}}=0, \cdots, \phi_{(N-1)\mathrm{c}}=0} }} 
\nonumber \\ & 
= 0 . 
\label{eqn:cond:condensate:Vonly}
\\
& \left. \frac{\partial^2}{\partial \phi_{s\mathrm{c}}^2} \Ueff(\{\phijc\}) \right|_{\{\phi_{0\mathrm{c}} = \phicVAC{0}, \phi_{1\mathrm{c}}=0, \cdots, \phi_{(N-1)\mathrm{c}}=0\}} 
\nonumber \\ &
=  \left. \frac{1}{V} \frac{\partial^2}{\partial \phi_{s\mathrm{c}}^2}  \int d\vec{x} \qexpectfree{\VpotHF} 
    \right|_{\substack{\{\phi_{0\mathrm{c}} = \phicVAC{0}, \phi_{1\mathrm{c}}=0, \cdots, \phi_{(N-1)\mathrm{c}}=0\}, \\ \{m_j=\massVAC{j}\} \hphantom{\phi_{1\mathrm{c}}=0, \cdots, \phi_{(N-1)\mathrm{c}}=0} }} 
\nonumber \\ &
= \massVAC{s}^2 .
\label{eqn:cond:mass:Vonly}
\end{align}
\end{subequations}

With $\qexpectfree{\varphi_j} = 0$, 
the normalized $q$-expectation value of $\VpotHF$ in the free particle approximation, $\qexpectfree{\VpotHF}$, is given by 
\begin{align}
\qexpectfree{\VpotHF} 
&= 
\frac{\lambda}{4} \Bigg\{
\left( \phic^2 + \qexpectfree{\varphi^2} - v^2 \right)^2 
   \nonumber \\ & \quad 
+ 4 \sum_{i=0}^{N-1} (\phiic)^2 \qexpectfree{\varphi_i^2}  
   \nonumber \\ & \quad 
+ 2 \sum_{i=0}^{N-1} \left( \qexpectfree{\varphi_i^2} \right)^2 
\Bigg\} 
- G \phi_{0\mathrm{c}} 
.
\end{align}

Here, we focus on the quantity $\qexpectfree{\varphi_j^2}$ and the removal of the divergent part.
The value $\qexpectfree{\varphi_j^2}$ is explicitly given in eq.~\eqref{varphi2:explicit}.
Equation~\eqref{varphi2:explicit} contains the divergent part: the first term of the right-hand side of eq.~\eqref{varphi2:explicit}.
In the present calculation,  we neglect the divergent terms in the following prescription, as done in the previous studies
\cite{Nicholas1999, Shu2005, Mao2006}.

We define the expectation value $\expectremoval{O}$:
\begin{align}
\expectremoval{O} := \qexpectfree{:O:} , 
\end{align}
where the $:O:$ represents the normal ordering with respect to the creation and annihilation operators with the mass $\massVAC{j}$.
We note that the removal term is the vacuum contribution which is the temperature-dependent, 
because $\massVAC{j}$ depends on the (physical) temperature.

In the present calculation, we need to remove the divergent part from the quantity $\qexpectfree{\varphi_j^2}$:
the quantity $\expectremoval{\varphi_j^2}$ is given by 
\begin{align}
\expectremoval{\varphi_j^2} = \qexpectfree{\varphi_j^2} - \frac{1}{4\pi^2 (\betaph)^2} F(x_j; -1, 0) . 
\end{align}
We replace $\qexpectfree{\varphi_j^2}$ with $\expectremoval{\varphi_j^2}$ in $\qexpectfree{\VpotHF}$, 
and denote the potential after the precedure as $\VpotHFR$.
We use $\VpotHFR$ instead of $\qexpectfree{\VpotHF}$: 
\begin{align}
\VpotHFR &= 
\frac{\lambda}{4} \Bigg\{
\left( \phic^2 + \expectremoval{\varphi^2} - v^2 \right)^2 
    \nonumber \\ & \quad 
+ 4 \sum_{i=0}^{N-1} (\phiic)^2 \expectremoval{\varphi_i^2}  
    \nonumber \\ & \quad 
+ 2 \sum_{i=0}^{N-1} \left( \expectremoval{\varphi_i^2} \right)^2 
\Bigg\} 
- G \phi_{0\mathrm{c}} .
\end{align}

To simplify eqs.~\eqref{eqn:cond:condensate:Vonly} and  \eqref{eqn:cond:mass:Vonly}
with the replacement from $\qexpectfree{\varphi_j^2}$ to  $\expectremoval{\varphi_j^2}$, 
we introduce the following notations.
\begin{subequations}
\begin{align}
K_j := \expectremoval{\varphi_j^2} ,\\
K := \sum_{i=0}^{N-1}  K_j . 
\end{align}
\end{subequations}
The quantity $K_j$ depends on the masses, but not the condensates. 
Equations~\eqref{eqn:cond:condensate:Vonly} and  \eqref{eqn:cond:mass:Vonly} are reduced to the following equations ($s \neq 0$):
\begin{subequations}
\begin{align}
&\lambda \left\{ (\phicVAC{0})^3 + (\overline{K} + 2\overline{K}_0 -v^2) \phicVAC{0} \right\} - G = 0,\\
&\phicVAC{s}  = 0,\\
&\lambda \left\{ 3 (\phicVAC{0})^2 + (\overline{K} + 2\overline{K}_0 -v^2)  \right\}  = \massVAC{0}^2,\\
&\lambda \left\{  (\phicVAC{0})^2 + (\overline{K} + 2\overline{K}_s -v^2)  \right\}  = \massVAC{s}^2, 
\end{align}
\end{subequations}
where $\overline{K}_j$ is $K_j$ with $\overline{m}_j$ and $\overline{K}$ is the sum of $\overline{K}_j$.
We finally obtain the equations with 
$\overline{K}_{\sigma} := \overline{K}_0$, $\overline{K}_{\pi} := \overline{K}_{s}$, $\massVAC{\sigma} := \massVAC{0}$, and $\massVAC{\pi} := \massVAC{s}$ $(s \neq 0)$:
\begin{subequations}
\begin{align}
& \Big(\frac{\phicVAC{0}}{v} \Big)^3+ \Big[ 3 \Big( \frac{\overline{K}_{\sigma}}{v^2} \Big) 
    \nonumber \\ & \qquad 
+ (N-1) \Big( \frac{\overline{K}_{\pi}}{v^2} \Big) - 1 \Big] \Big( \frac{\phicVAC{0}}{v}  \Big)
- \frac{G}{\lambda v^3} = 0,
\label{eqn:third}
\\
&\lambda 
\Big\{ 3 \Big(\frac{\phicVAC{0}}{v} \Big)^2+ \Big[ 3 \Big( \frac{\overline{K}_{\sigma}}{v^2} \Big) 
    \nonumber \\ & \qquad 
+ (N-1) \Big( \frac{\overline{K}_{\pi}}{v^2} \Big) - 1 \Big] \Big\}  
= \Big( \frac{\massVAC{\sigma}}{v} \Big)^2 ,
\label{eqn:gap:m:sigma}
\\
&\lambda 
\Big\{ \Big(\frac{\phicVAC{0}}{v} \Big)^2+ \Big[ \Big( \frac{\overline{K}_{\sigma}}{v^2} \Big) 
    \nonumber \\ & \qquad 
+ (N+1) \Big( \frac{\overline{K}_{\pi}}{v^2} \Big) - 1 \Big] \Big\}  
= \Big( \frac{\massVAC{\pi}}{v} \Big)^2 , 
\label{eqn:gap:m:pi}
\end{align}
\end{subequations}
where $\phicVAC{s}$ is zero for $s \neq 0$.

Equation~\eqref{eqn:third} as a function of $(\phicVAC{0}/v)$ can be solved. 
We represent eq.~\eqref{eqn:third} as follows:
\begin{align}
(\phicVAC{0}/v)^3 + 3 \eta (\phicVAC{0}/v) + 2\kappa = 0, 
\end{align}
\begin{subequations}
where $\kappa$ is negative. 
The real number solution for $\eta \ge - \sqrt[3]{|\kappa|^2}$ is represented as

\begin{align}
(\phicVAC{0}/v) = \sqrt[3]{-\kappa + \sqrt{\kappa^2 + \eta^3}} + \sqrt[3]{-\kappa - \sqrt{\kappa^2 + \eta^3}} , 
\label{sol1}
\end{align}
where $\sqrt[3]{-1} = -1$.
The real number solution for $\eta < - \sqrt[3]{|\kappa|^2}$ is represented as

\begin{align}
& (\phicVAC{0}/v) =  2 \sqrt{-\eta} \cos(\theta/3), 
\label{sol2}
\\ & \quad  
\tan\theta = \frac{\sqrt{-(\kappa^2+\eta^3)}}{-\kappa}\quad (0 < \theta < \pi/2) .
\nonumber
\end{align}
The solution represented by eq.~\eqref{sol1} and eq.~\eqref{sol2} is continuous at $\eta = - \sqrt[3]{|\kappa|^2}$.
\label{solution}
\end{subequations}

In the next section, we solve the equations, eqs.~\eqref{eqn:gap:m:sigma} and \eqref{eqn:gap:m:pi},  numerically 
with eq.~\eqref{solution}. 


\section{Numerical solutions of gap equations}
\label{Sec:NumericalCalculation}

In this section, we attempt to calculate the condensate, the sigma mass, and the pion mass as a function of the physical temperature $\Tph$
for various $q$ by solving eqs.~\eqref{eqn:gap:m:sigma} and \eqref{eqn:gap:m:pi} with eq.~\eqref{solution}.  

We set the parameters of the linear sigma model to generate the pion mass, sigma mass,  and pion decay constant:
the pion mass is 135 MeV, the sigma mass is 600 MeV, and the pion decay constant is 92.5 MeV. 
The number of the field $N$ is set to four. 
The values of the parameters, $\lambda$, $v$, and $G$, are 
approximately $20$, $87.4$ MeV, and $(119\ \mathrm{MeV})^3$, respectively. 
These values generate the above masses and decay constant.

\begin{figure}
\centering
  \subfigure{\includegraphics[width=0.45\textwidth]{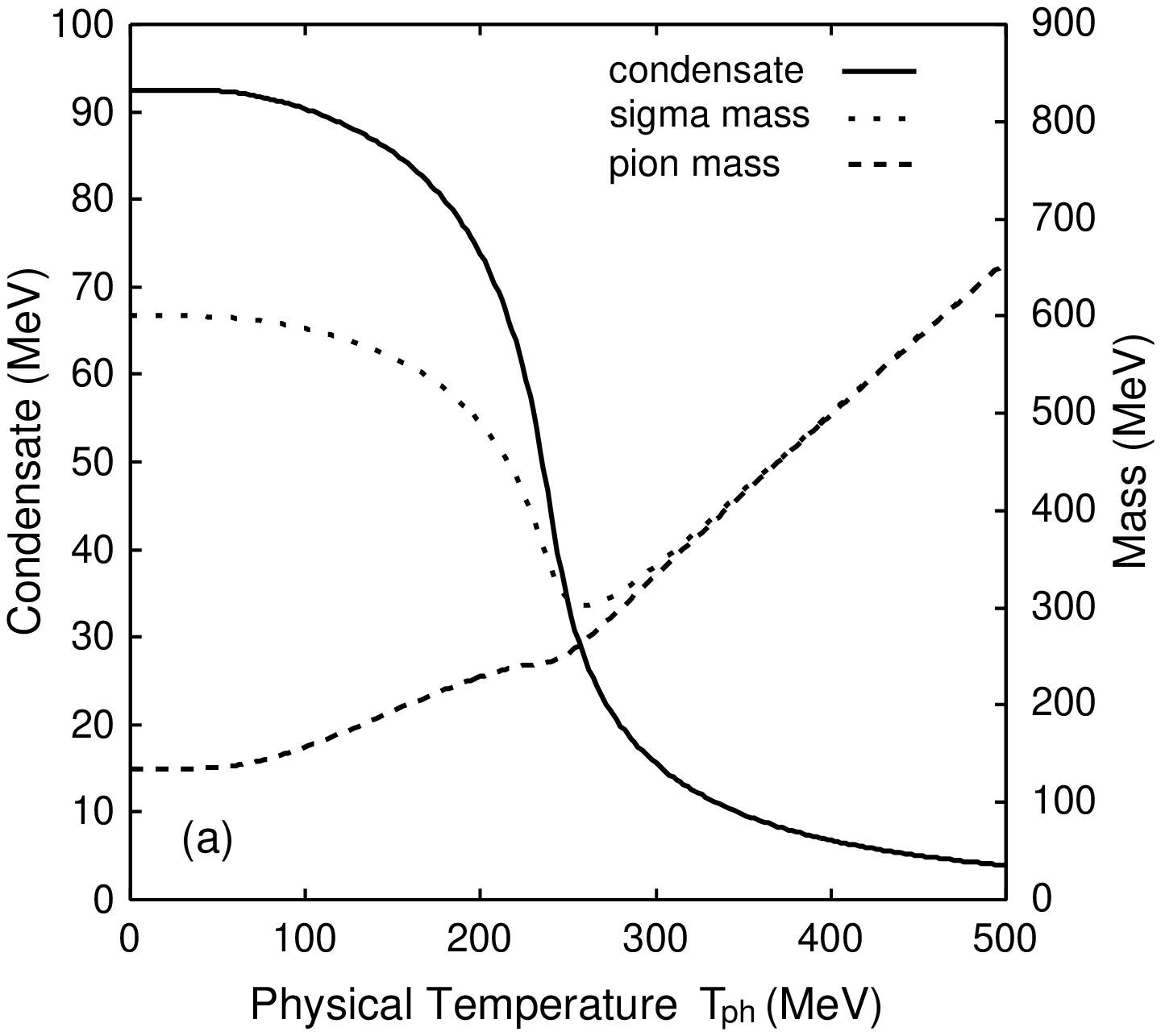}\label{Fig:q0.9}}
  \subfigure{\includegraphics[width=0.45\textwidth]{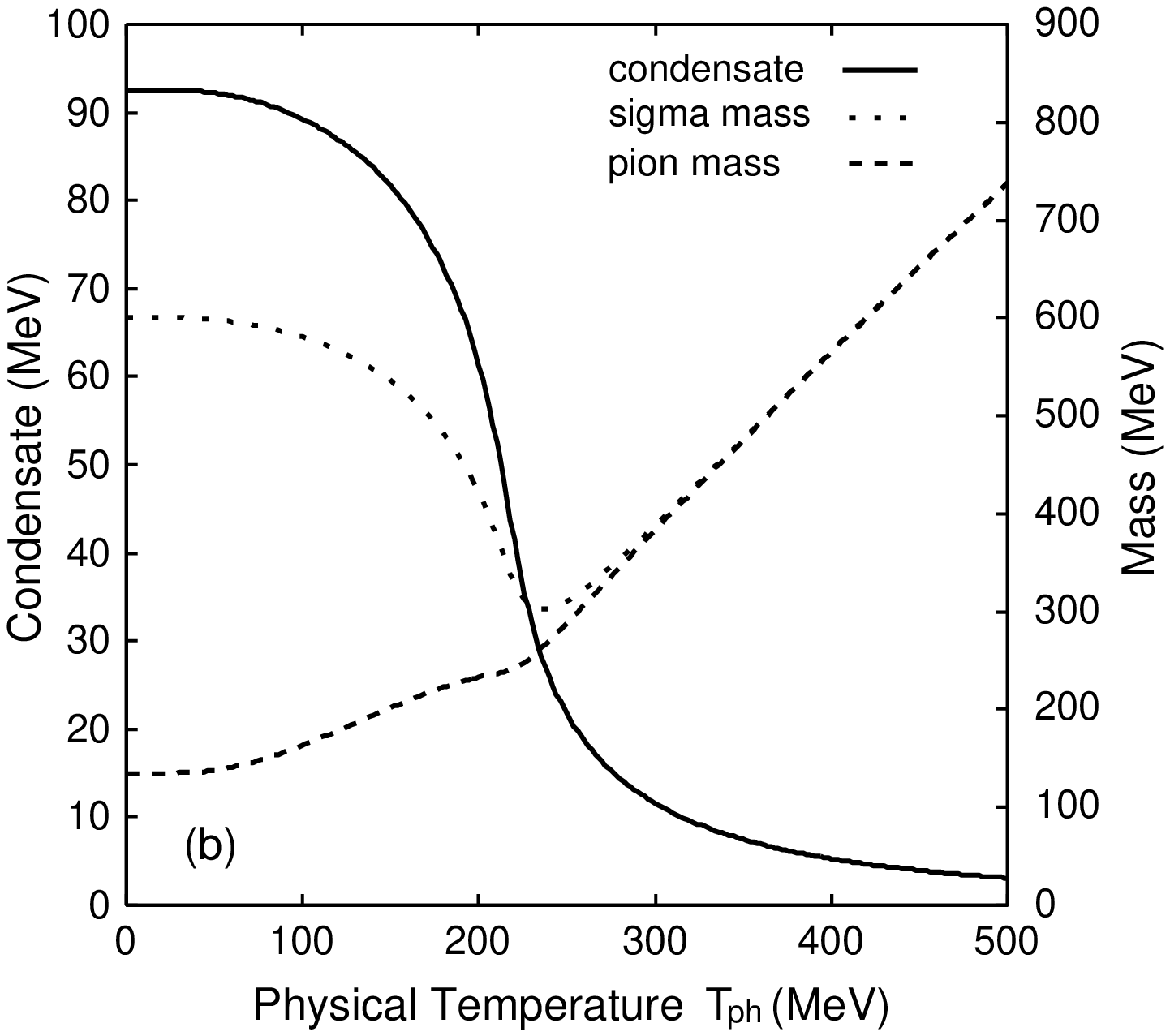}\label{Fig:q1.0}}
  \subfigure{\includegraphics[width=0.45\textwidth]{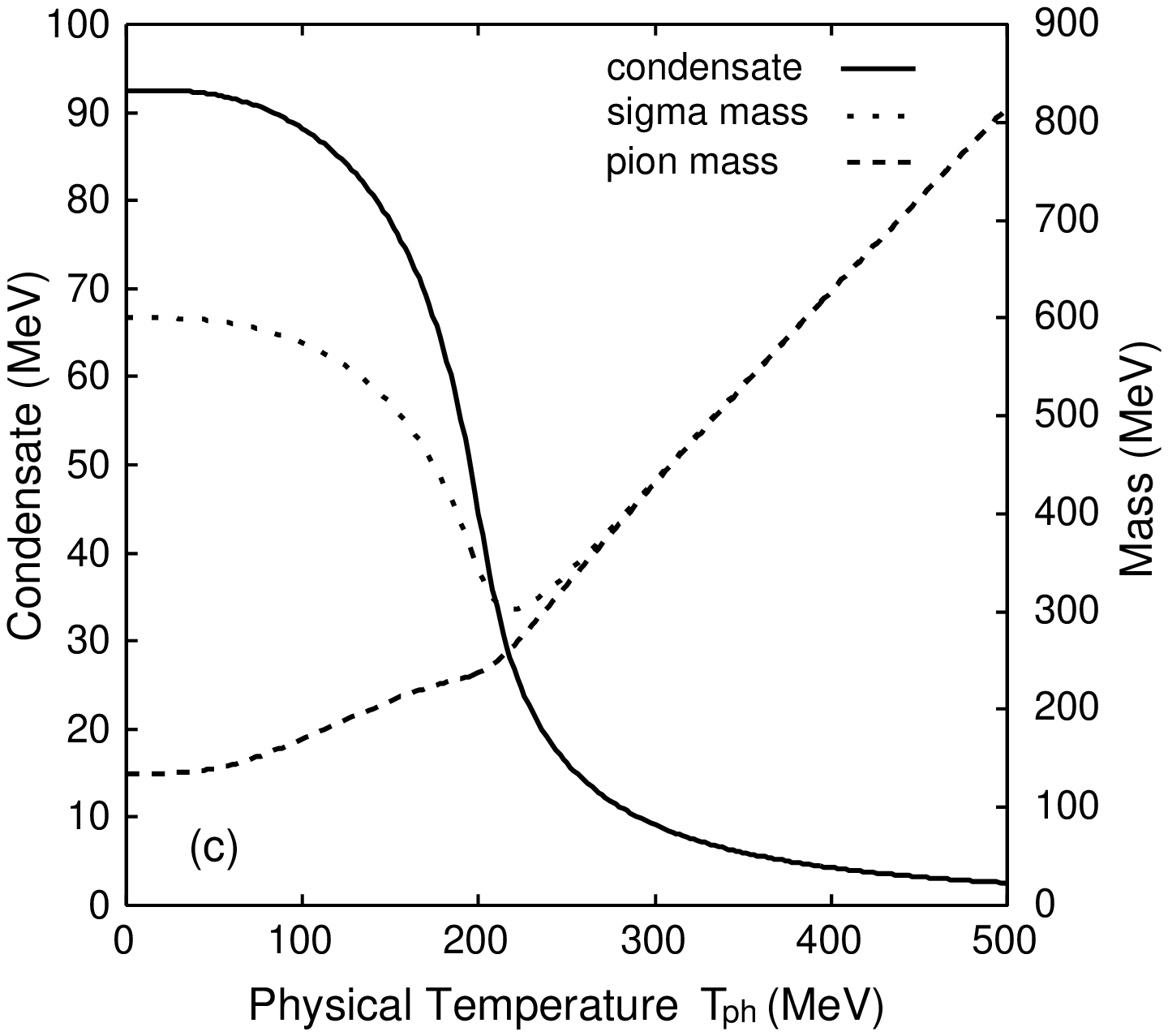}\label{Fig:q1.1}}
  \caption{Physical temperature dependences of the condensate, sigma mass, and pion mass for (a) $q=0.9$, (b) $1.0$, and (c) $1.1$.}
\end{figure}

Figures~\ref{Fig:q0.9}, \ref{Fig:q1.0}, and \ref{Fig:q1.1} show 
the physical temperature dependences of the condensate, the sigma mass, and the pion mass
for $q=0.9$, $1.0$, and $1.1$, respectively.
The numerical results at $q=1.0$ correspond to the results in the Boltzmann-Gibbs statistics. 
The behaviors shown in fig.~\ref{Fig:q1.0} are similar to those shown in the previous work \cite{Nicholas1999}.
As shown in figs.~\ref{Fig:q0.9}, \ref{Fig:q1.0}, and \ref{Fig:q1.1}, 
the behavior of the condensate at $q \neq  1.0$ is similar to that at $q=1.0$ (BG statistics). 
These resemblances are also seen for the sigma mass and pion mass.

\begin{figure}
\begin{center}
\includegraphics[width=0.45\textwidth]{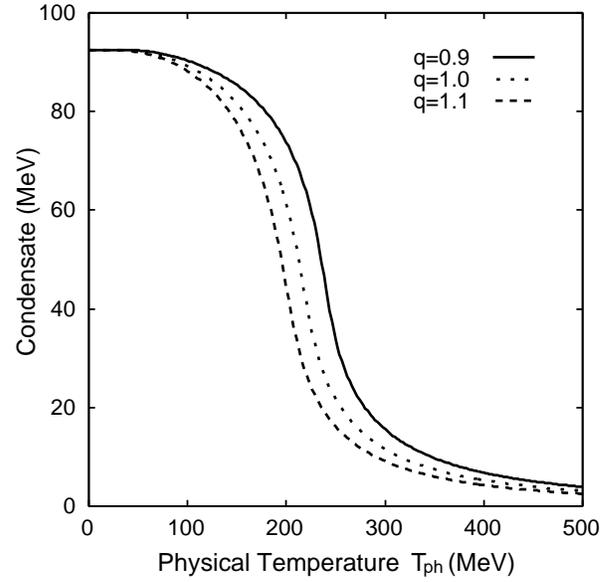}
\end{center}
\caption{Physical temperature dependence of the condensate for $q=0.9$, $1.0$, and $1.1$.}
\label{Fig:condensate}
\end{figure}

Figure~\ref{Fig:condensate} shows the physical temperature dependences of the condensate for $q=0.9$, $1.0$, and $1.1$.
The curves are similar, and the condensate at $q$ is smaller than that at $q'$ for $q>q'$.
Therefore, the chiral symmetry restoration occurs at low physical temperature for large $q$, 
while the restoration occurs at high physical temperature for small $q$.

\begin{figure}
\centering
  \subfigure{\includegraphics[width=0.45\textwidth]{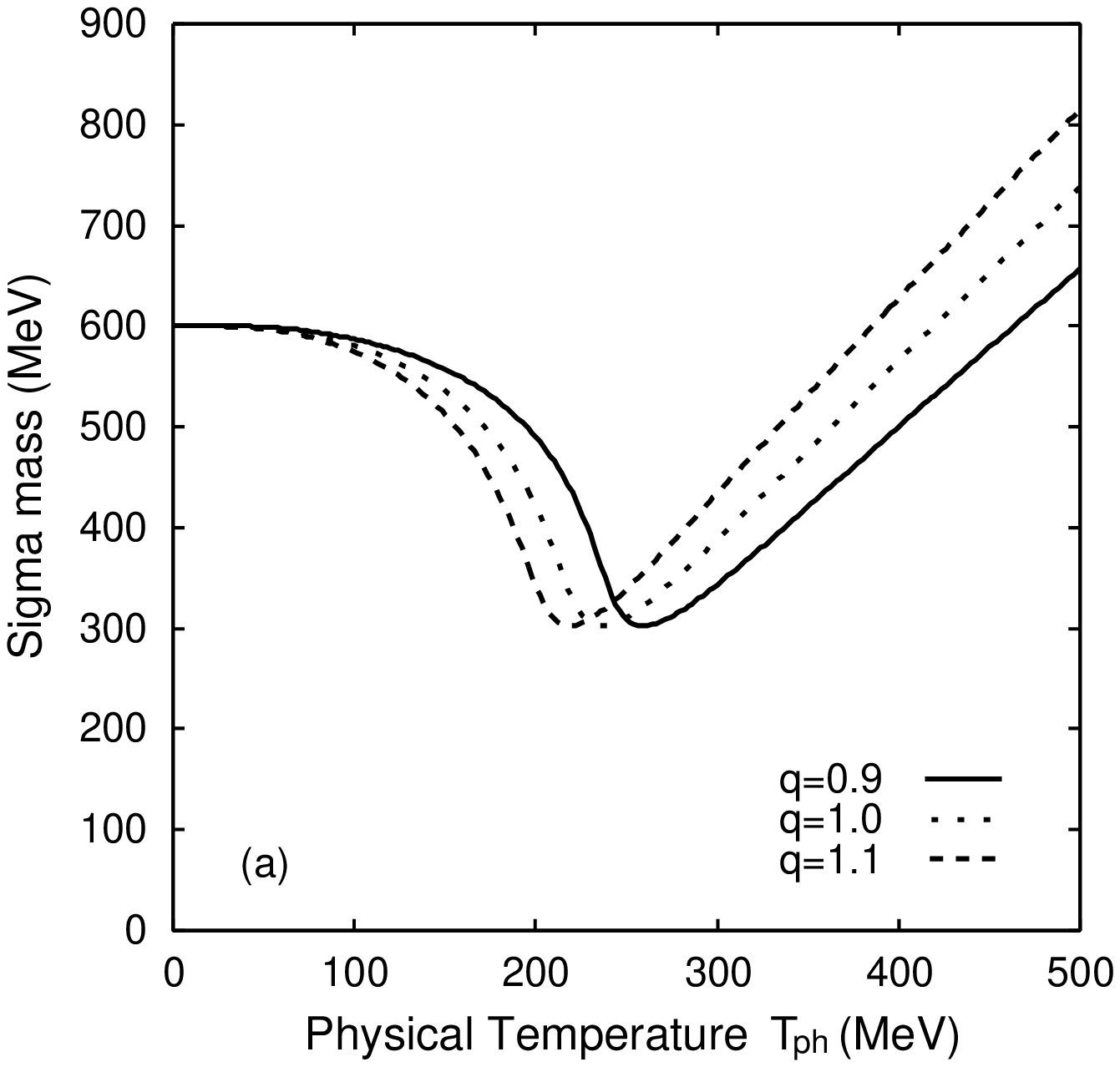}\label{Fig:sigma}}
  \subfigure{\includegraphics[width=0.45\textwidth]{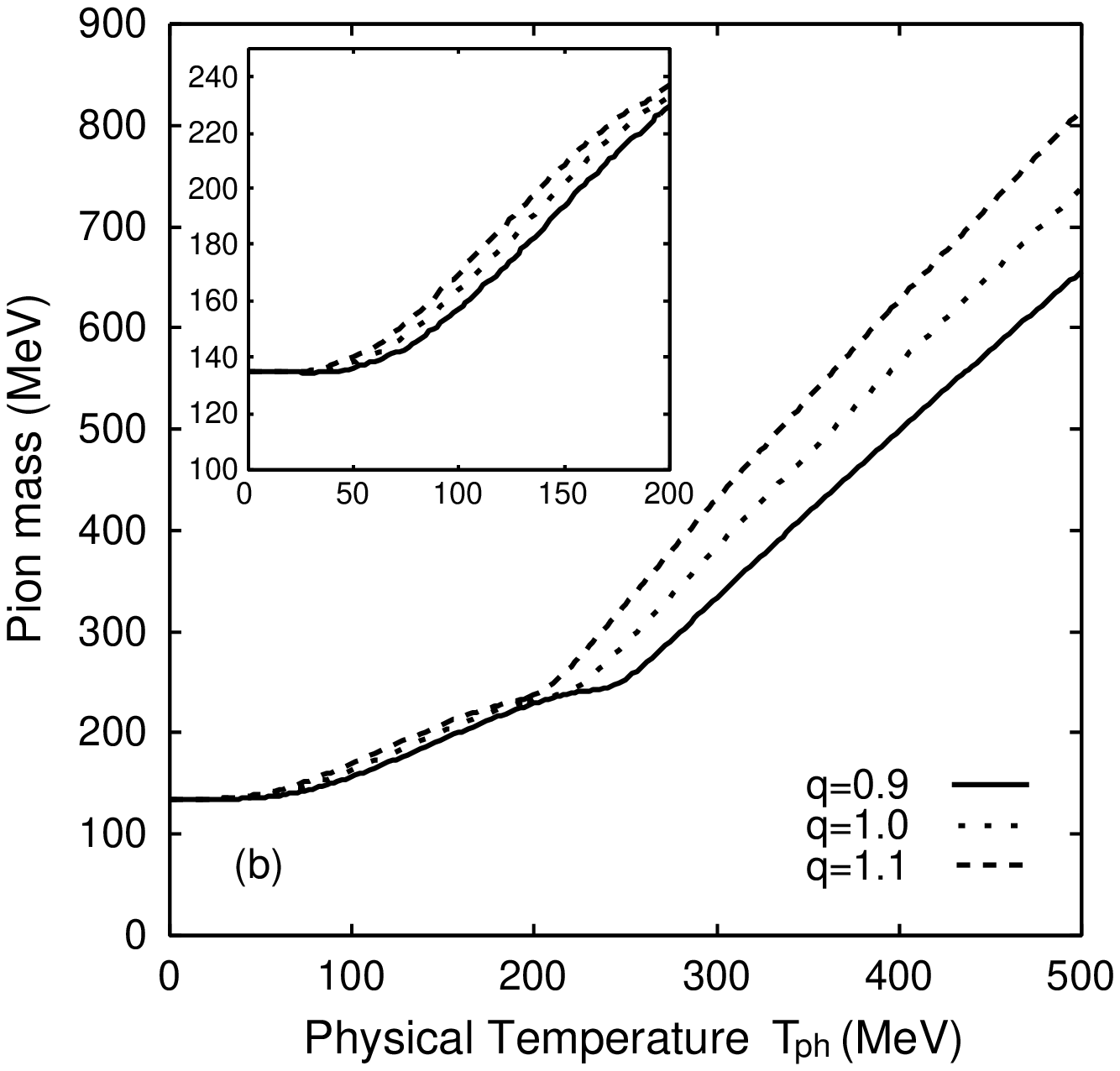}\label{Fig:pion}}
  \caption{Physical temperature dependences of (a) the sigma mass and (b) the pion mass for $q=0.9$, $1.0$, and $1.1$.}
\end{figure}

Figure~\ref{Fig:sigma} shows the physical temperature dependences of the sigma mass for $q=0.9$, $1.0$, and $1.1$.
The sigma mass decreases, reaches minimum, and increases after that,  as  the physical temperature increases. 
The sigma mass at $q$ is lighter than that at $q'$ for $q>q'$ at low physical temperature, 
while the sigma mass at $q$ is heavier than that at $q'$ for $q>q'$ at high physical temperature.
This behavior of the sigma mass is explained by the behavior of the condensate. 
The decrease of the sigma mass at low physical temperature for large $q$ implies 
that chiral symmetry restoration occurs at low physical temperature for large $q$.

Figure~\ref{Fig:pion} shows the physical temperature dependences of the pion mass for $q=0.9$, $1.0$, and $1.1$.
As shown in the figure, 
the pion mass increases monotonically with the physical temperature, and 
the pion mass  at $q$ is heavier than that at $q'$ for $q>q'$. 
The difference betweeen the mass at $q$ and the mass at $q' (\neq q)$  is small for $\Tph < 200$ MeV, 
while the difference grows with the physical temperature for $\Tph > 200$ MeV.
This implies that the effects of the Tsallis nonextensive statistics on the pion mass are weak at low physical temperature.


\section{Discussion and conclusion}
\label{Sec:Dicussion}
We investigated the physical temperature $\Tph$ dependences of the condensate, the sigma mass, and the pion mass  
in the Tsallis nonextensive statistics of the entropic parameter $q$, 
when the deviation from the Boltzmann-Gibbs statistics, $|1-q|$, is small. 
We applied the Hartree factorization to the Hamiltonian and free particle approximation 
to the normalized $q$-expectation values of the squares of the fields. 
The divergent terms of the normalized $q$-expectation values were dropped. 
We solved the self-consistent equations numerically, 
and obtained the physical temperature dependences of the condensates, the sigma mass, and the pion mass for $q=0.9$, $1.0$, and $1.1$.

We found that the condensate at $q$ is smaller than that at $q'$ for $q>q'$. 
The sigma mass and the pion mass reflect the value of the condensate. 
The sigma mass at $q$ is lighter than that at $q'$ for $q>q'$ at low physical temperature, 
while the mass at $q$ is heavier than that at $q'$ for $q>q'$ at high physical temperature. 
The pion mass is a monotonically increasing function of the physical temperature, 
and the pion mass at $q$ is heavier than that at $q'$ for $q>q'$.

These behaviors should be explained by the fact that the tail of the distribution becomes elongated with $q$. 
The effects of the normalized $q$-expectation value of the square of the field is large when the tail of the distribution is long.
Therefore, the modification caused by the normalized $q$-expectation value is large for large $q$.
This indicates that the chiral symmetry restoration occurs at low physical temperature for large $q$.

The effects of the Tsallis nonextensive statistics of small $|1-q|$ on the pion mass are small 
for $\Tph \le 200$ MeV, as shown in the calculations in the present study.  
It is not easy to find the effects of the statistics on the pion mass at low physical temperature. 
In other words, at low physical temperature, the pion mass at $q\neq 1$ is close to the mass at $q=1$. 
The effects on the pion mass may be found at high physical temperature. 
In contrast, the effects on the sigma mass appear even for $\Tph \le 200$ MeV.

In summary, we studied the effects of the Tsallis nonextensive statistics of small $|1-q|$ on the condensate, 
the sigma mass, and the pion mass. 
The condensate and the sigma mass are affected by the statistics.  
The pion mass is also affected by the statistics except for $\Tph \le 200$ MeV. 
That is, the effects of the statistics on the pion mass is small for $\Tph \le 200$ MeV.

We hope that this work is helpful for the readers to study the phase transition 
within the framework of the Tsallis nonextensive statistics.

\appendix
\section{Hartree Factorization}

In this appendix, we summarize the Hartree factorization \cite{Boyanovsky:PRD55, Boyanovsky:PRD56} for this paper to be self-contained. 
The basic idea is to represent the operators in quadratic expression.

\subsection{Single Field}
We begin with a single field $\varphi$ with the constraint $\<\varphi>=0$.  
The operator $\varphi^{2n}$ is approximated by the following form.  
\begin{align}
\varphi^{2n} \sim A (\<\varphi^2>)^{n-1} \varphi^2 + B (\<\varphi^2>)^{n-1} \<\varphi> \varphi + C (\<\varphi^2>)^{n} .
\label{varphi2n:app}
\end{align}
The second term vanishes because of the constraint $\<\varphi>=0$.  
We focus on the coefficients $A$ and $C$. 

The coefficient $A$ is obtained by counting the combination of $\varphi$. 
The number of pairs constructed from $\varphi^{2n}$  is given by ${}_{2n} C_2$. 
It is possible to do this procedure recursively, we find the combination of $(n-1)$ pairs
\begin{equation}
{}_{2n}C_2  \times {}_{2n-2}C_2 \times \cdots \times {}_{4}C_2 = \frac{(2n)(2n-1)\cdots 3 }{2^{n-1}}  = \frac{(2n)!}{2^n} .
\end{equation}
This is over-counting, and we must divide the value by the number of the orders of $(n-1)$ pairs. 
Therefore, we get 
\begin{align}
A = \frac{(2n)!}{2^n (n-1)! } .
\end{align}
Next, we approximate $\<\varphi^{2n}>$ by $E \<\varphi^2>^n$.
The coefficient $E$ is obtained in the same manner:
\begin{align}
E = \frac{(2n)!}{2^n n! } .
\end{align}

We obtain the coefficient $C$ by taking the expectation value of Eq.~\eqref{varphi2n:app}.
This gives 
\begin{align}
\< \varphi^{2n}> \sim E  \<\varphi^2>^n = (A+C)   \<\varphi^2>^n .  
\end{align}
The coefficient $C$ is 
\begin{align}
C = - \frac{(2n)! (n-1)}{2^n n!} .
\end{align}

The Hartree factorization of $\varphi^{2n}$ is given by
\begin{align}
\varphi^{2n}  &\sim 
\frac{(2n)!}{2^n (n-1)!} (\<\varphi^2>)^{n-1} \varphi^2 
\nonumber \\ & \quad 
- \frac{(2n)! (n-1)}{2^n n!} (\<\varphi^2>)^{n}  
\qquad (n \ge 1).  
\end{align}
For example, the Hartree factorization of $\varphi^4$ is given by 
\begin{align}
\varphi^{4} \sim 6 \<\varphi^2> \varphi^2 -  3 (\<\varphi^2>)^{2} .  
\end{align}

In the same way,  we obtain the Hartree factorization of $\varphi^{2n+1}$: 
\begin{align}
\varphi^{2n+1} \sim \frac{(2n+1)!}{2^n n!} (\<\varphi^2>)^n \varphi  \qquad (n \ge 1). 
\end{align}
This gives the Hartree factorization of $\varphi^3$:
\begin{align}
\varphi^{3} \sim 3 \<\varphi^2> \varphi .
\end{align}

\subsection{Two Fields}
Next, we treat two fields, $\varphi$ and $\psi$, with $\<\varphi>=0$, $\<\psi>=0$, and $\<\varphi \psi>=0$. 
The Hartree factorization of $\varphi^{2m} \psi^{2n}$  and  $\varphi^{2m} \psi^{2n+1}$ are derived in the similar way.

The Hartree factorization of  $\varphi^{2m} \psi^{2n}$ is given by the following form. 
\begin{align}
\varphi^{2m} \psi^{2n} \sim &
\varphi^{2} \mathrm{\ terms} +  \psi^{2} \mathrm{\ terms} + \varphi \psi \mathrm{\ terms} 
\nonumber \\ & 
+ \<\varphi^2>^{m} \<\psi^2>^{n} \mathrm{\ terms} .
\label{eqn:hartree:2m:2n}
\end{align}
The coefficients of $\varphi^2$, $\psi^2$, and $\varphi \psi$ are constructed from $\<\varphi^2>$, $\<\psi^2>$, and $\<\varphi \psi>$.
For example, 
the first term of eq.~\eqref{eqn:hartree:2m:2n} may contain several terms such as \\
$\<\varphi^2>^{m-2} \<\psi^2>^{n-1} \<\varphi \psi>^2 \varphi^{2}$. 
However, the term \\
$\<\varphi^2>^{m-2} \<\psi^2>^{n-1} \<\varphi \psi>^2 \varphi^{2}$ vanishes because of the assumption $\<\varphi \psi>=0$.
The remaining form of the first term is  $\<\varphi^2>^{m-1} \<\psi^2>^{n}\varphi^{2}$. 
Therefore, the Hartree factorization under the present assumptions is given by 
\begin{align}
\varphi^{2m} \psi^{2n} &\sim 
    A (\<\varphi^2>)^{m-1} (\<\psi^{2}>)^{n}   \varphi^{2} 
\nonumber \\ & \quad 
+ B (\<\varphi^2>)^{m} (\<\psi^{2}>)^{n-1} \psi^{2} 
+ C (\<\varphi^2>)^{m} (\<\psi^2>)^{n}  . 
\label{eqn:hartree:2m:2n:compact}
\end{align}

The coefficient $A$ in eq.~\eqref{eqn:hartree:2m:2n:compact}  is obtained in the similar way. 
The coefficient $A$ is given by 
\begin{align} 
A = \frac{(2m)!}{2^m (m-1)!} \times  \frac{(2n)!}{2^n n!} . 
\end{align} 
The coefficient $B$ in eq.~\eqref{eqn:hartree:2m:2n:compact}  is given by 
\begin{align} 
B = \frac{(2m)!}{2^m m!} \times  \frac{(2n)!}{2^n (n-1)!} .
\end{align} 
The term $\<\varphi^{2m} \psi^{2n}>$ is approximated as $E (\<\varphi^{2}>)^m (\<\psi^{2}>)^n$. 
The coefficient $E$ is given by 
\begin{align} 
E = \frac{(2m)!}{2^m m!} \times  \frac{(2n)!}{2^n n!} .
\end{align} 
We obtain the coefficient $C$ by taking the average of eq.~\eqref{eqn:hartree:2m:2n:compact}.
We get the equation
\begin{align}
E = A+B+C. 
\end{align}
The final representation of the Hartree factorization of $\varphi^{2m} \psi^{2n}$ with $\<\varphi>=\<\psi>=\<\varphi \psi>=0$ is 
\begin{align}
\varphi^{2m} \psi^{2n} &\sim 
   \left( \frac{(2m)! }{2^m (m-1)!} \frac{(2n)!}{2^n n!} \right)  (\<\varphi^2>)^{m-1} (\<\psi^{2}>)^{n}   \varphi^{2} 
\nonumber \\ & \quad
+ \left( \frac{(2m)!}{2^m m!} \frac{(2n)!}{2^n (n-1)!} \right) (\<\varphi^2>)^{m} (\<\psi^{2}>)^{n-1} \psi^{2} 
\nonumber \\& \quad
- (m+n-1) \left( \frac{(2m)!}{2^m m!} \frac{(2n)!}{2^n n!} \right) (\<\varphi^2>)^{m} (\<\psi^2>)^{n}  . 
\label{eqn:hartree:2m:2n:final}
\end{align}
The Hartree factorization of $\varphi^2 \psi^2$ is given with Eq.~\eqref{eqn:hartree:2m:2n:final} with $m=n=1$. 
The factorization is 
\begin{align}
\varphi^2 \psi^2 \sim 
\varphi^2 \< \psi^2 > + \< \varphi^2 > \psi^2  - \< \varphi^2> \< \psi^2> .
\label{HF:m=n=1}
\end{align}

We obtain the Hartree factorization of $\varphi^{2m} \psi^{2n+1}$, as in the case of $\varphi^{2m} \psi^{2n}$. 
The Hartree factorization of $\varphi^{2m} \psi^{2n+1}$ with the assumptions $\<\varphi>=\<\psi>=\<\varphi \psi>=0$ is given by
\begin{align}
\varphi^{2m} \psi^{2n+1} \sim \frac{(2m)!}{2^m m!} \frac{(2n+1)!}{2^n n!}  (\<\varphi^2>)^m  (\<\psi^2>)^n  \psi .
\end{align}
This result gives the factorization of $\varphi^2 \psi$:
\begin{align}
\varphi^2 \psi \sim \< \varphi^2 > \psi .
\end{align}

The above discussion will be extended to multi-fields.



\begin{thebibliography}{99}

\bibitem{Book:Tsallis} \mbox{C.~Tsallis}, 
{\it Introduction to Nonextensive Statistical Mechanics} (Springer Science+Business Media, LLC, 2010). 


\bibitem{Tsallis1998} C.~Tsallis, R.~S.~Mendes, and A.~R.~Plastino, 
``The role of constraints within generalized nonextensive statistics'', 
Physica A \textbf{261},  534  (1998).


\bibitem{Kalyana2000} S.~Kalyana Rama, 
``Tsallis statistics: averages and a physical interpretation of the Lagrange multiplier $\beta$'',
Phys.~Lett.~A \textbf{276},  103  (2000). 

\bibitem{Abe-PLA2001}  S.~Abe, S.~Martinez,  F.~Pennini, and A.~Plastino, 
``Nonextensive thermodynamics relations'', 
Phys.~Lett.~A~\textbf{281}, 126  (2001). 

\bibitem{Aragao-PhysicaA2003} H.~H.~Arag\~ao-R\^ego, D.~J.~Soares, L.~S.~Lucena, L.~R.~da~Silva, E.~K.~Lenzi, and Kwok~Sau~Fa, 
``Bose-Einstein and Fermi-Dirac distributions in nonextensive Tsallis Statistics: an exact study'', 
 Physica A \textbf{317},  199  (2003) . 


\bibitem{Eicke-prepri2003} E.~Ruthotto,  
``Physical temperature and the meaning of the $q$ parameter in Tsallis statistics'',  
arXiv:cond-mat/0310413.

\bibitem{Toral-PhysicaA2003} R.~Toral,  
``On the definition of physical temperature and pressure for nonextensive thermodynamics'',
Physica A \textbf{317},  209  (2003). 

\bibitem{Suyari-PTPsupple2006} H.~Suyari, 
``The Unique Non Self-Referential q-Canonical Distribution and the Physical Temperature Derived from the Maximum Entropy Principle in Tsallis Statistics'', 
Prog. of Theor. Phys. supplement \textbf{162},  79  (2006).



\bibitem{Alberico2009} W.~M.~Alberico, A.~Lavagno,
``Non-extensive statistical effects in high-energy collisions'',
Eur.~Phys.~J.~A \textbf{40},  313  (2009).


\bibitem{Urmossy2011-PLB701}  
K.~Urmossy, G.~G.~Barnaf\"oldi, and T.~S.~Bir\'o,
 ``Generalized Tsallis Statistics in Electron-Positron Collisions'', 
Phys.~Lett.~B \textbf{701},  111  (2011).




\bibitem{Cleymans2012} J.~Cleymans and D.~Worku, 
``The Tsallis distribution in proton-proton collisions at $\sqrt{s}=0.9$ TeV at the LHC'',
J.~Phys.~G: Nucl. Part. Phys. \textbf{39},  025006  (2012). 


\bibitem{Cleymans2013-PLB723} 
J.~Cleymans, G.~I.~Lykasov, A.~S.~Parvan, A.~S.~Sorin, O.~V.~Teryaev, and  D.~Worku, 
``Systematic properties of the Tsallis Distribution: Energy Dependence of Parameters in High-Energy $p$-$p$ Collisions'', 
Phys.~Lett.~B \textbf{723},  351 (2013).


\bibitem{Marques2015} L.~Marques, J.~Cleymans, and A.~Deppman, 
``Description of high-energy $pp$ collisions using Tsallis thermodynamics: Transverse momentum and rapidity distributions'', 
Phys.~Rev.~D \textbf{91},  054025 (2015).

\bibitem{GS2015} L.~McLerran and M.~Praszalowicz, 
``Geometrical scaling and the dependence of the average transverse momentum on the multiplicity and energy for the ALICE experiment'',
Phys.~Lett.~B \textbf{741}, 246  (2015). 


\bibitem{Azmi2015} M.~D.~Azmi and J.~Cleymans, 
``The Tsallis distribution at large transverse momenta'', 
Eur.~Phys.~J.~C \textbf{75},  430  (2015).  


\bibitem{Zheng2016}  H.~Zheng and L.~Zhu,
``Comparing the Tsallis Distribution with and without Thermodynamical Description in $p+p$ Collisions'',
Adv.~in~High~Ener.~Phys. \textbf{2016}, 9632126 (2016). 


\bibitem{Thakur-AHEP2016}
D.~Thakur, S.~Tripathy, P.~Garg, R.~Sahoo, and J.~Cleymans, 
``Indication of a Differential Freeze-Out in Proton-Proton and Heavy-Ion Collisions at RHIC and LHC Energies'', 
Adv.~in~High~Ener.~Phys. \textbf{2016}, 4149352  (2016) .



\bibitem{Lao2017} H.-L.~Lao, F.-H.~Liu, and R.~A.~Lacey,
``Extracting kinetic freeze-out temperature and radial flow velocity from an improved Tsallis distribution'',
Eur.~Phys.~J.~A \textbf{53}, 44  (2017).


\bibitem{Cleymans2017-WoC} J.~Cleymans, M.~D.~Azmi, A.~S.~Parvan, and O.~V.~Teryaev, 
``The Parameters of the Tsallis Distribution at the LHC'', 
EPJ Web of Conferences \textbf{137}, 11004  (2017).


\bibitem{Cleymans2017} A.~Khuntia, S.~Tripathy, R.~Sahoo, and J.~Cleymans, 
``Multiplicity dependence of non-extensive parameters for strange and multi-strange particles in proton-proton collisions at $\sqrt{s}=7$ TeV at the LHC'', 
Eur.~Phys.~J.~A \textbf{53}, 103 (2017) . 


\bibitem{Yin2017} X.~Yin, L.~Zhu, and H.~Zheng, 
``A new two-component model for hadron production in heavy-ion collisions'', 
Adv.~in~High~Ener.~Phys. \textbf{2017}, 6708581 (2017).

\bibitem{Osada-Ishihara-2018} T.~Osada and M.~Ishihara, 
``Event-by-event mean $p_{T}$ fluctuations and transverse size of color flux tube generated in $p$-$p$ collisions at $\sqrt{s}=0.90$ TeV'', 
J.~Phys.~G: Nucl.~Part.~Phys. \textbf{45}, 015104 (2018).


\bibitem{Bhattacharyya2017-JPhysG45} T.~Bhattacharyya, J.~Cleymans, L.~Marques, S.~Mogliacci, and M.~W.~Paradza, 
``On the precise determination of the Tsallis parameters in proton-proton collisions at LHC energies'', 
J.~Phys.~G:~Nucl.~Part.~Phys. \textbf{45},  055001 (2018). 



\bibitem{Si-AHEP2018}  Rui-Fang~Si, Hui-Ling~Li, and Fu-Hu~Liu,  
``Comparing Standard Distribution and Its Tsallis Form of Transverse Momenta in High Energy Collisions'', 
Adv.~in~High~Ener.~Phys.  \textbf{2018}, 7895967 (2018).


\bibitem{Shen2018-PhysicaA492} 
K.~M.~Shen, T.~S.~Bir\'o, E.~K.~Wang, 
``Different Non-extensive Models for heavy-ion collisions'', 
Physica A \textbf{492}, 2353  (2018).


\bibitem{Osada2019-Prepri}
T.~Osada and T.~Kumaoka, 
``Saturation momentum scale extracted from semi-inclusive transverse spectra in high-energy pp collisions'', 
arXiv:1904.10823v1. 
 



\bibitem{Ishihara2017-1} M.~Ishihara, 
``Momentum distribution and correlation due to mass difference caused by power-like distribution'', 
Int. J. Mod. Phys. E \textbf{26}, 1750039 (2017).


\bibitem{Ishihara2017-2} M. Ishihara, 
``Transverse momentum fluctuation under the Tsallis distribution at high energies'', 
Int. J. Mod. Phys. E \textbf{26}, 1750071 (2017).

\bibitem{Ishihara2018} M.~Ishihara, 
``Momentum distribution and correlation for a free scalar field in the Tsallis nonextensive statistics based on density operator'', 
Eur.~Phys.~J.~A \textbf{54}, 164 (2018).




\bibitem{Rozynek2009} J.~Ro\.zynek and G.~Wilk,  
``Nonextensive effects in the {Nambu-Jona-Lasinio} model of {QCD}'',  
J.~Phys.~G: Nucl.~Part.~Phys. \textbf{36}, 125108 (2009).

\bibitem{Ishihara2015} M. Ishihara, 
``Effects of the Tsallis distribution in the linear sigma model'', 
Int. J. Mod. Phys. E \textbf{24}, 1550085 (2015).

\bibitem{Rozynek2016} J.~Ro\.zynek and G.~Wilk, 
``Nonextensive {Nambu-Jona-Lasinio} Model of QCD matter'', 
Eur.~Phys.~J.~A \textbf{52}, 13 (2016).  

\bibitem{Ishihara2016} M. Ishihara, 
``Chiral phase transitions in the linear sigma model in the Tsallis nonextensive statistics'', 
Int. J. Mod. Phys. E \textbf{25}, 1650066 (2016).


\bibitem{Shen2017} K.-M.~Shen, H.~Zhang, D.-F.~Hou, B.-W.~Zhang, and E.-K.~Wang,
``Chiral phase transition in linear sigma model with non-extensive statistical mechanics'',
Adv.~in~High~Ener.~Phys. \textbf{2017}, 4135329 (2017). 

\bibitem{Ishihara2019} M.~Ishihara, 
``Chiral phase transition within the linear sigma model in the Tsallis nonextensive statistics based on density operator'', 
Int. J. Mod. Phys. E \textbf{28}, 1950020 (2019).  


\bibitem{Boyanovsky:PRD55} D.~Boyanovsky, D.~Cormier,  H.~J.~de~Vega, and R.~Holmann, 
``Out of equilibrium dynamics of an inflationary phase transition'', 
Phys.~Rev.~D \textbf{55},  3373 (1997).


\bibitem{Boyanovsky:PRD56} D.~Boyanovsky, D.~Cormier,  H.~J.~de~Vega, R.~Holmann, A.~Singh, and M.~Srednichi,
``Scalar field dynamics in Friedmann-Robertson-Walker spacetimes'', 
Phys.~Rev.~D \textbf{56}, 1939 (1997).


\bibitem{Nicholas1999} N.~Petropoulos,  
``Linear sigma model and chiral symmetry at finite temperature'', 
J.~Phys.~G: Nucl. Part. Phys. \textbf{25}, 2225 (1999). 


\bibitem{Shu2005} S.~Shu and  J.-R. Li, 
``Bose-Einstein condensation and chiral phase transition in linear sigma model'',
J.~Phys.~G: Nucl. Part. Phys. \textbf{31}, 459 (2005).  


\bibitem{Mao2006} H.~Mao, N.~Petropoulos, S.~Shu, and W.-Q. Zhao, 
``The linear sigma model at a finite isospin chemical potential'', 
J.~Phys.~G: Nucl. Part. Phys. \textbf{32}, 2187 (2006).  

\end{thebibliography}
\end{document}